# Effect of stepwise adjustment of Damping factor upon PageRank

Subhajit Sahu[1], Kishore Kothapalli[1], Dip Sankar Banerjee[2]
[1]International Institute of Information Technology, Hyderabad
[2]Indian Institute of Technology, Jodhpur

**Abstract** — The effect of adjusting damping *factor α*, from a *small initial value $α_0$* to the *final desired $α_f$ value*, upon then iterations needed for PageRank computation is observed. Adjustment of the *damping factor* is done in *one or more steps*. Results show *no improvement* in performance over a *fixed damping factor* based PageRank.

## 1. Introduction

Web graphs are by default reducible, and thus the **convergence rate** of the *power-iteration method* is the rate at which $α^k → 0$, where *α* is the *damping factor*, and *k* is the *iteration count*. An estimate of the number of iterations needed to converge to a *tolerance τ* is $log_{10} τ / log_{10} α$ [1]. For $τ = 10^{-6}$ and *α = 0.85*, it can take roughly *85 iterations* to converge. For *α = 0.95*, and *α = 0.75*, with the same *tolerance $τ = 10^{-6}$*, it takes roughly *269* and *48 iterations* respectively. For $τ = 10^{-9}$, and $τ = 10^{-3}$, with the same *damping factor α = 0.85*, it takes roughly *128* and *43 iterations* respectively. Thus, adjusting the *damping factor* or the *tolerance parameters* of the PageRank algorithm can have a significant effect on the *convergence rate*.

## 2. Method

The idea behind this experiment was to *adjust the damping factor α in steps*, to see if it might help *reduce* PageRank computation time. The PageRank computation first starts with a *small initial damping factor α = $α_0$*. After the ranks have *converged*, the *damping factor α* is updated to the next damping factor step, say $α_1$ and PageRank computation is continued again. This is done until the *final desired value* of $α_f$ is reached. For example, the computation starts initially with $α = α_0 = 0.5$, lets ranks converge quickly, and then switches to $α = α_f = 0.85$ and continues PageRank computation until it converges. This **single-step** change is attempted with the initial (fast converge) *damping factor* $α_0$ from *0.1* to *0.8*. Similar to this, two-step, three-step, and four-step changes are also attempted. With a **two-step** approach, a *midpoint* between the initial damping value $α_0$ and $α_f = 0.85$ is selected as well for the second set of iterations. Similarly, **three-step** and **four-step** approaches use *two* and *three* midpoints respectively.

## 3. Experimental setup

A *small sample graph* is used in this experiment, which is stored in the *MatrixMarket (.mtx)* file format. The experiment is implemented in Node.js, and executed on a personal laptop. Only the *iteration count* of each test case is measured. The *tolerance* $τ = 10^{-5}$ is used for all test cases. Statistics of each test case is printed to *standard output (stdout)*, and redirected to a *log file*, which is then processed with a *script* to generate a *CSV file*, with each row representing the details of a *single test case*. This *CSV file* is imported into *Google Sheets*, and necessary tables are set up with the help of the *FILTER* function to create the *charts*.

## 4. Results

From the results, as shown in figure 4.1, it is clear that **modifying the damping factor α in *steps* is not a good idea**. The *standard fixed damping factor PageRank*, with $α = 0.85$, converges in *35 iterations*. Using a *single-step* approach increases the total number of iterations required, by at least *4 iterations* (with an initial *damping factor* $α_0 = 0.1$). Increasing $α_0$ further *increases* the total number of iterations needed for computation. Switching to a *multi-step* approach also increases the number of iterations needed for convergence. The two-step, three-step, and four-step approaches require a total of atleast *49*, *60*, and *71 iterations* respectively. Again, increasing $α_0$ continues to *increase* the total number of iterations needed for computation. A possible explanation for this effect is that the ranks for different values of the *damping factor α* are *significantly different*, and switching to a different damping factor α after each step mostly leads to *recomputation*.

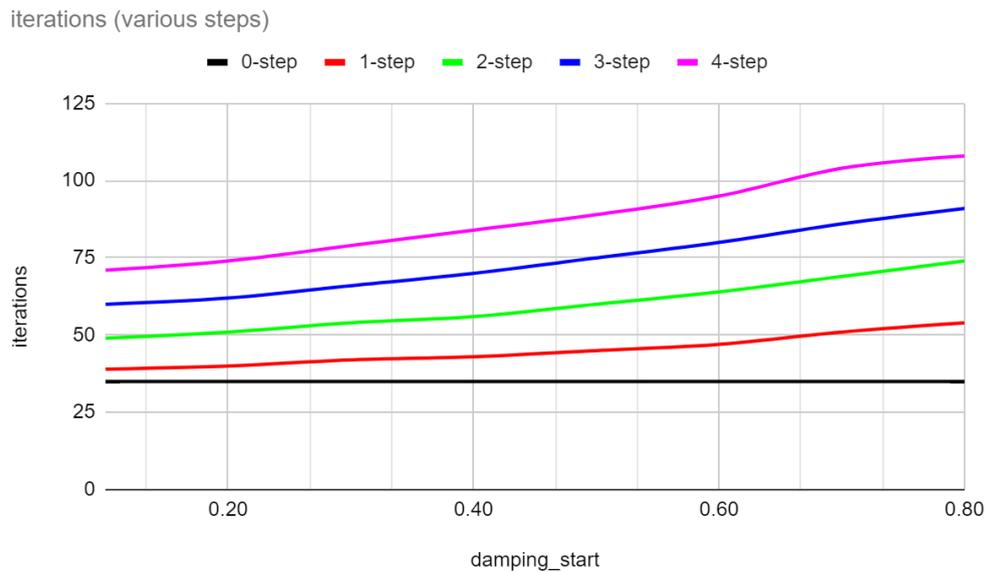

Figure 4.1: Iterations required for PageRank computation, when *damping factor α* is adjusted in *1-4 steps*, starting with an *initial small damping factor $α_0$ (damping_start)*. *0-step* is the *fixed damping factor PageRank*, with *α = 0.85*.

## 5. Conclusion

Adjusting the *damping factor α* in *steps* is not a good idea, most likely because ranks obtained for even *nearby damping factors* are significantly different. Fixed damping factor PageRank continues to be the best approach. The links to source code, along with data sheets and charts, for adjusting damping factor in steps [2] is included in references.